\documentclass[pra, twocolumn, floatfix, superscriptaddress, longbibliography, preprintnumbers]{revtex4-1}

\usepackage{amsmath, amssymb, bbm, bbold}
\usepackage{graphicx}
\usepackage{color}

\newcommand{\ket}[1]{{| #1 \rangle}}
\newcommand{\bra}[1]{{\langle #1 |}}
\newcommand{\ee}{{\rm e}}
\newcommand{\ii}{{\rm i}}
\newcommand{\dd}{{\rm d}}
\newcommand{\Tr}{{\rm Tr}}
\newcommand{\sinc}{{\rm sinc}}

\begin{document}

\title{Quantum evolution in disordered transport}

\author{Clemens Gneiting}
\email{clemens.gneiting@riken.jp}
\affiliation{Quantum Condensed Matter Research Group, RIKEN, Wako-shi, Saitama 351-0198, Japan}
\author{Franco Nori}
\email{fnori@riken.jp}
\affiliation{Quantum Condensed Matter Research Group, RIKEN, Wako-shi, Saitama 351-0198, Japan}
\affiliation{Department of Physics, University of Michigan, Ann Arbor, Michigan 48109-1040, USA}

\date{\today}

\begin{abstract}
We analyze the propagation of quantum states in the presence of weak disorder. In particular, we investigate the reliable transmittance of quantum states, as potential carriers of quantum information, through disorder-perturbed waveguides. We quantify wave-packet distortion, backscattering, and disorder-induced dephasing, which all act detrimentally on transport, and identify conditions for reliable transmission. Our analysis relies on the treatment of the nonequilibrium dynamics of ensemble-averaged quantum states in terms of quantum master equations.
\end{abstract}

\preprint{\textsf{published in Phys.~Rev.~A~{\bf 96}, 022135 (2017)}}

\maketitle

\section{Introduction}

Quantum transport theory has mainly been concerned with the impact of irregular perturbations (i.e.,~disorder) on classical, macroscopic observables, such as the conductivity of a material. This follows  a clear technological motivation, since disorder can have severe consequences, possibly hindering transport up to the complete trapping of charge carriers in the wire, i.e.~localization, turning a metal into an insulator \cite{Lifshits1988introduction, Rammer1991quantum, Beenakker1997random, Akkermans2007mesoscopic, Nishiguchi1993phonon, Lin1996analytical}. In these considerations, the mere arrival of a charge carrier at the output end of a specimen is of relevance, so details of the precise transmitted quantum state are usually omitted.

The situation is changing with the ongoing maturing of quantum technologies. Quantum particles, along with their quantum behavior, are not confined anymore to serve only classical technology purposes; they are now expected to fulfill genuinely quantum tasks, assuming their roles as carriers of quantum information. In a future quantum computer, for instance, a bus may transport qubits between different locations \cite{Friesen2007efficient, Cronin2009optics, Buluta2011natural, Huang2013spin, Zhao2016doppler, Lekitsche2017blueprint}. In quantum communication, on the other hand, photons may deliver quantum information encoded in their optical angular momentum \cite{Mair2001entanglement}. Another field of relevance is the fundamental testing of quantum mechanics with matter wave interferometry \cite{Brezger2002matter, Cronin2009optics, Khakimov2016ghost}. Applications are numerous and growing.

\begin{figure}[htb]
	\includegraphics[width=0.90\columnwidth]{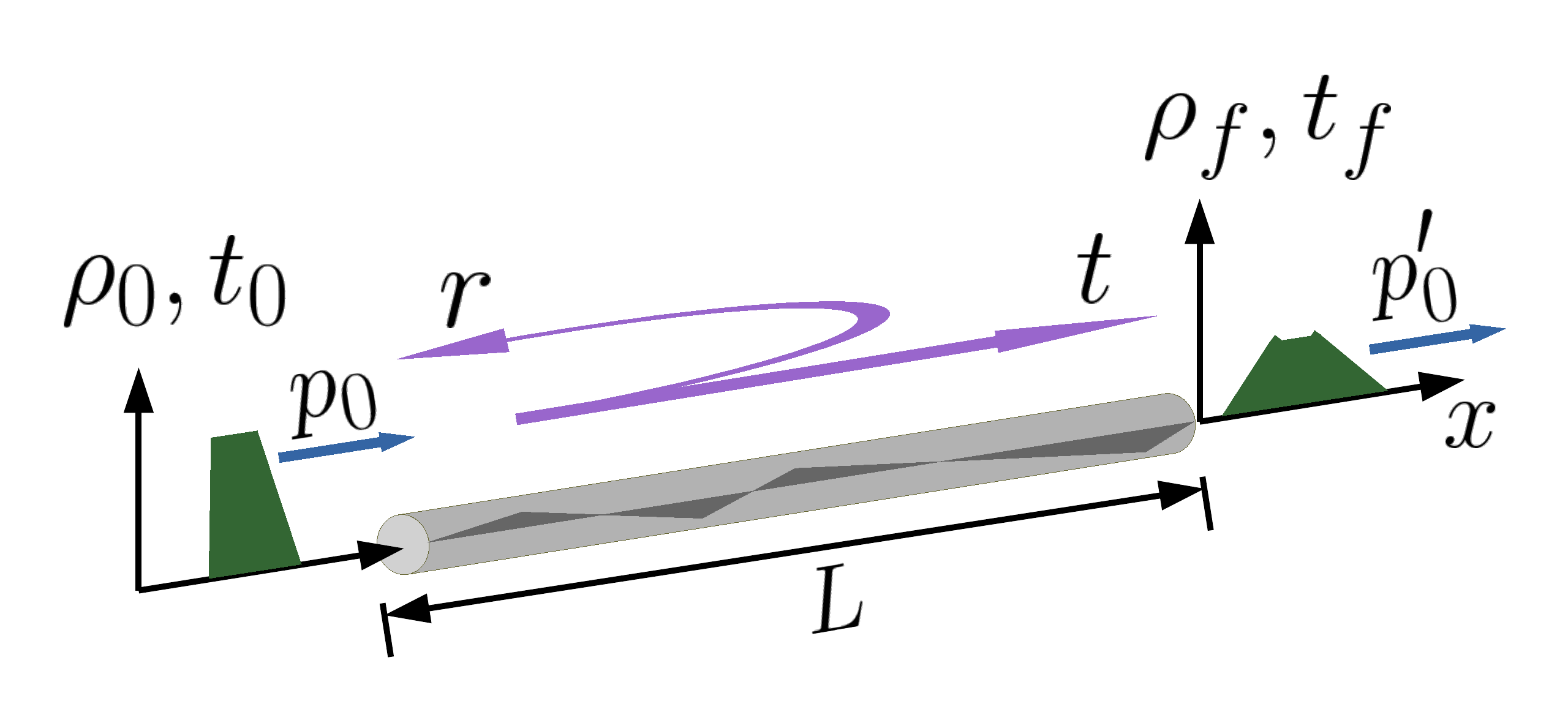}
	\caption{\label{Fig:disordered_transport_channel} (Color online) Disordered transport channel for quantum states. A particle of mass $m$ and momentum $p_0$, described by the initial state $\rho_0$, propagates through an imperfect waveguide. The disorder potential is symbolized by the darker zigzag inlay. While traditional transport theory has been focusing on conduction properties (indicated by $t$ and $r$), reducing quantum particles to charge carriers, quantum technologies, with quantum states carrying quantum information, ask for a more refined description, encompassing comprehensive disorder modifications on the level of both, the populations and coherences of $\rho_f$.}
\end{figure}

Even the highly controlled environments of quantum devices are not devoid of uncontrolled perturbations in the form of disorder. Think, for instance, of stray fields or surface effects. But the relevant question is not, whether transmission takes place, but rather with which precision. Even if the transmission loss is negligible, disorder-induced dephasing can still spoil the performance of the device; similarly, state distortion can undermine the necessary fidelity. In Figure~\ref{Fig:disordered_transport_channel} we sketch this refined transport task. To address these issues, we need a detailed and comprehensive description of (small) nonequilibrium deviations from a given input.

In this article, we develop such a description. To this end, we take up the recently introduced treatment of disordered systems in terms of quantum master equations \cite{Gneiting2016incoherent, Kropf2016effective}. The latter have proven to be well suited to describing the transient dynamics of disorder-averaged systems. In \cite{Gneiting2016incoherent}, a quantum master equation for general disorder configurations was provided, valid, however, only in the limit of short times. In \cite{Kropf2016effective}, exact quantum master equations for specific, symmetric disorder configurations were derived, excluding, however, the rich and important family of transport problems. Here, we fill this gap and substantially extend the quantum master equation approach, introducing coupled disorder channels, to perturbative treatments of arbitrary disorder configurations beyond short times. By virtue of this method, we comprehensively describe the disorder-perturbed evolution of propagating, massive quantum particles, and evaluate it for wave packet distortion, backscattering, and disorder-induced dephasing.

\section{Disorder-perturbed propagation}

A general disorder ensemble is comprised of a set of Hamiltonians $\hat{H}_{\varepsilon}$ occurring with probability $p_{\varepsilon}$, where the (multi-)index $\varepsilon$ labels the different disorder realizations (for simplicity, we write integrals throughout, e.g., $\int \dd \varepsilon \, p_{\varepsilon} = 1$). In our case, single realizations $\rho_{\varepsilon}$ follow unitary dynamics, $\partial_t \rho_{\varepsilon} = -(\ii/\hbar) [\hat{H}_{\varepsilon},\rho_{\varepsilon}]$, formally solved by $\rho_{\varepsilon}(t) = \hat{U}_{\varepsilon}(t) \rho_0 \hat{U}_{\varepsilon}^{\dagger}(t)$, with the time evolution operator $\hat{U}_{\varepsilon}(t) = \exp[-(\ii/\hbar) \hat{H}_{\varepsilon} t]$ and $\rho_0$ an arbitrary initial state (identical for all realizations). The ensemble-averaged state $\overline{\rho}(t) \equiv \int \dd \varepsilon \, p_{\varepsilon} \, \rho_{\varepsilon}(t)$ is then obtained from $\overline{\rho}(t) = \int \dd \varepsilon \, p_{\varepsilon} \, \hat{U}_{\varepsilon}(t) \rho_0 \hat{U}_{\varepsilon}^{\dagger}(t)$.

We analyze the propagation of a single particle of mass $m$ and (mean) momentum $p_0$ in one dimension (1D), subject to a homogeneous disorder potential but otherwise free, described by the Hamiltonian $\hat{H}_{\varepsilon} = \frac{\hat{p}^2}{2 m} + V_{\varepsilon}(\hat{x})$. The disorder potential may exhibit two-point correlations $C(x-x') \equiv \int \dd \varepsilon \, p_{\varepsilon} \, V_{\varepsilon}(x) V_{\varepsilon}(x')$ and may vanish on average, $\int \dd \varepsilon \, p_{\varepsilon} \, V_{\varepsilon}(x) = 0$. We assume that the disorder potential is weak, in the sense that the kinetic energy $\frac{p_0^2}{2 m}$ of the particle is large compared to its magnitude; otherwise any wave packet would rapidly be strongly distorted and trapped, rendering reliable transport meaningless.

Below, we introduce the coupled disorder channel equations (\ref{Eq:coupled_disorder_channels}), and, as their byproduct, the perturbative quantum master equation (\ref{Eq:perturbative_master_equation}). The latter allows us to derive analytic solutions for the evolution of ensemble-averaged quantum states. In the case considered here, such solution is best represented in terms of the (complex-valued) characteristic function $\overline{\chi}_t(s,q) = \int \dd x \dd p \, \ee^{-\frac{\ii}{\hbar}(q x-p s)} \overline{W}_t(x,p)$ of the (real-valued) phase-space representation (for a brief introduction, see, for instance, \cite{Schleich2011quantum} or \cite{Gneiting2013quantum}) $\overline{W}_t(x,p) = \frac{1}{2 \pi \hbar} \int \dd x' \ee^{\frac{\ii}{\hbar} p x'} \bra{x-\frac{x'}{2}} \overline{\rho}_t \ket{x+\frac{x'}{2}}$ of the ensemble-averaged state $\overline{\rho}_t$.

Evaluating the quantum master equation (\ref{Eq:perturbative_master_equation}) for the ensemble-averaged propagation of a wave packet with momentum $p_0$ and exposed to a weak disorder potential, we obtain the phase-space solution
\begin{subequations} \label{Eq:phase-space_solution}
\begin{align}
\overline{\chi}_t(s,q) = \chi_0 \left( s-\frac{q}{m} t, q \right) \exp \left[ -\overline{F}_t(s,q) \right],
\end{align}
where the disorder influence is summarized by
\begin{widetext}
	\begin{align} \label{Eq:disorder_influence}
	\overline{F}_t(s,q) = -\frac{2}{\hbar^2} \int \dd q' G(q') \int_0^t \dd t_1 \int_0^{t_1} \dd t_2 \left\{ \cos \left[ \frac{q' (q'+2 p_0) t_2}{2 m \hbar} \right] \ee^{-\ii \frac{q' q t_2}{2 m \hbar}} \ee^{\frac{\ii}{\hbar} q' (s-\frac{q}{m} [t-t_1])} - \cos \left[ \frac{q' (q'-2 p_0) t_2}{2 m \hbar} \right] \ee^{\ii \frac{q' q t_2}{2 m \hbar}} \right\} .
	\end{align}
\end{widetext}
\end{subequations}
This is our main result and the starting point of our analysis. The propagation equation (\ref{Eq:phase-space_solution}) describes the averaged evolution of the full disorder-perturbed quantum state. Note that the propagation equation~(\ref{Eq:phase-space_solution}) is derived from the quantum master equation~(\ref{Eq:perturbative_master_equation}) without further approximations and therefore is well-defined in the entire phase space.

We emphasize that the propagation equation~(\ref{Eq:phase-space_solution}) holds for arbitrary, sufficiently well-behaved initial states $\chi_0(s,q)$ with momentum width $\sigma_p \ll p_0$, where $\chi_0 \left( s-\frac{q}{m} t, q \right)$ describes the free, undisturbed propagation of the initial state; this includes, e.g., spatially delocalized superposition states. If evaluated, we hereafter assume, to be generic, pure, Gaussian initial states, $\psi_0(x) = \exp \left[-\frac{1}{4}(\frac{x}{\sigma})^2 + \frac{\ii}{\hbar} p_0 x \right]/[\sqrt{2 \pi} \sigma]^{1/2}$, with $\hbar/\sigma \ll p_0$. The characteristics of the disorder enter through the momentum transfer distribution $G(q) = \frac{1}{2 \pi \hbar} \int \dd x \, \ee^{-\frac{\ii}{\hbar} q x} C(x)$ (see also \cite{Gneiting2016incoherent}). Note that for $C(-x)=C(x)$ we obtain $G(-q)=G(q)$, which we assume throughout. While the solution (\ref{Eq:phase-space_solution}) holds for general two-point correlations $C(x)$, we assume, to be generic, Gaussian correlations for evaluations, $C(x) = C_0 \, \exp[-(x/\ell)^2]$, with the correlation length $\ell$. This then gives rise to a Gaussian momentum-transfer distribution,
\begin{align} \label{Eq:Gaussian_momentum_transfer_distribution}
G(q) = \frac{C_0 \ell}{2 \sqrt{\pi} \hbar} \exp \left[ -\frac{1}{4} \left( \frac{q \ell}{\hbar} \right)^2 \right] .
\end{align}
The momentum scale $\hbar/\ell$ will be essential to characterizing the disorder-induced dynamics. Note that the strength of the disorder potential is contained in $C_0$.

\section{Backscattering}

Backscattering opposes reliable quantum state transmission and should, if possible, be strongly suppressed. While classically not present in the limit considered (of large kinetic energy), quantum mechanically, backscattering can happen at any energy; correspondingly, in 1D, Anderson localization may occur for arbitrarily weak disorder. With disorder being the only source of backscattering here, scattering is elastic.

If we evaluate the momentum distribution $P_t(p) = \bra{p} \overline{\rho}_t \ket{p} = \frac{1}{2 \pi \hbar} \int \dd s \, \ee^{-\frac{\ii}{\hbar} p s} \overline{\chi}_t(s,0)$ for the propagation equation~(\ref{Eq:phase-space_solution}) and for times $\frac{p_0}{m} t \gg \ell,\sigma$, we can approximate (without loss of generality $p_0>0$)
\begin{align} \label{Eq:backscattering_momentum_distribution}
P_t(p) = P_0(p) + \frac{2 \pi m t}{p_0 \hbar} G(2 p_0) \left[ P_0(p+2p_0) - P_0(p) \right] ,
\end{align}
where $P_0(p)$ denotes the momentum distribution of the (unspecified) initial state. Equation~(\ref{Eq:backscattering_momentum_distribution}) describes, within the approximation, the linear-in-time build-up of a backscattering peak at $-p_0$. However, in view of Eq.~(\ref{Eq:Gaussian_momentum_transfer_distribution}), we find that the backscattering rate can, given appropriate correlations, be exponentially suppressed, and thus, depending on the transport task, for all practical purposes be omitted, if $p_0 \ell \, \hbar^{-1} \gg 1$, or, in terms of the de~Broglie wavelength $\lambdabar_{\rm dB} = \hbar/p_0$ of the particle, $\lambdabar_{\rm dB} \ll \ell$. This condition thus constitutes a benchmark requirement for reliable quantum state transmittance. Let us remark that similar results for backscattering rates can be obtained with different methods \cite{Izrailev2012anomalous}.

Note that reproducing backscattering proves the quantum nature of our theory, beyond semiclassical approximation. Moreover, Eq.~(\ref{Eq:backscattering_momentum_distribution}) can, similar to its stationary counterparts \cite{Izrailev2012anomalous}, explain the emergence of effective mobility edges when tailoring momentum transfer distributions with sharp cut-offs (corresponding to long-range correlations). Evaluated for periodic disorder, Eq.~(\ref{Eq:backscattering_momentum_distribution}) reflects backscattering-free propagation (off-resonant) and Bragg scattering (on-resonant), respectively.

Propagation equation~(\ref{Eq:phase-space_solution}) correctly describes the onset of the first backscattering event, while two or more scattering processes are not included. These become relevant, when the single-backscattering peak has grown significantly in size, opening the door to a second scattering event. While this restriction limits the temporal validity of Eq.~(\ref{Eq:phase-space_solution}), it is already beyond the considered regime of reliable transport, where already single-scattering events are detrimental and the single-backscattering peak should be small if not negligible.

\section{Decorrelation regime}

Hereafter, we assume that $p_0 \gg \hbar/\ell$ is fulfilled and that backscattering is, to first order, negligible. As we now show, even then the disorder potential may have a significant impact on the evolution of the quantum state.

To see this, we evaluate the moments of momentum, $\langle \hat{p}^n \rangle\!(t) = (-\ii \hbar)^n \frac{\partial^n \chi_t}{\partial s^n}(s,0)|_{s=0}$. Specifically, we focus on the mean momentum $\langle \hat{p} \rangle\!(t)$ and the momentum variance $\langle (\Delta \hat{p})^2 \rangle\!(t)$. In the case of free, undisturbed propagation, the momentum distribution is time independent and thus $\langle \hat{p} \rangle\!(t) = p_0$ and, in the case of the above Gaussian initial state, $\langle (\Delta \hat{p})^2 \rangle\!(t) = \hbar^2/(4 \sigma^2)$.

The mean momentum in the presence of a disorder potential, as one deduces from the propagation equation~(\ref{Eq:phase-space_solution}), is described by $\langle \hat{p} \rangle\!(t) = p_0 + \frac{t^2}{\hbar^2} \int \dd q' G(q') q' \sinc^2 \left[ \frac{q' t}{4 m \hbar} (q'+2 p_0) \right]$. Evaluating this to $\mathcal{O}(\frac{\hbar}{p_0 \ell})$, we obtain, with Eq.~(\ref{Eq:Gaussian_momentum_transfer_distribution}), $\langle \hat{p} \rangle\!(t) = p_0 - \frac{2 m^2 C_0}{p_0^3} \left\{ 1 - \left[ 1 + \left( \frac{p_0 t}{m \ell} \right)^2 \right] \ee^{-\left( \frac{p_0 t}{m \ell} \right)^2} \right\}$. We thus find that, in the course of a decorrelation period, the ensemble-averaged mean momentum undergoes a shift, which, after $\frac{p_0}{m} t \gg \ell$, takes the plateau value
\begin{align} \label{Eq:momentum_shift}
\langle \hat{p} \rangle\!(t \gg \ell m/p_0 ) = p_0 - \frac{2 m^2 C_0}{p_0^3} .
\end{align}
This shift, which has also a classical counterpart, must be attributed to the fact that positive and negative potential variations affect the momentum differently, giving rise to an asymmetric distortion of the wave packet in momentum.

We define the onset of backscattering dominance as the time $t_{\rm bd}$ at which the reduction of the mean momentum due to backscattering exceeded the decorrelation momentum shift. Since backscattering affects the mean momentum as $\langle \hat{p} \rangle_{\rm bs}\!(t) = p_0 - 4 \pi m G(2 p_0) \hbar^{-1} t$, we obtain $t_{\rm bd} = \hbar m C_0 [2 \pi p_0^3 G(2 p_0)]^{-1}$.

The momentum shift is accompanied by a broadening of the momentum variance, which, when $\frac{p_0}{m} t \gg \ell$, assumes the plateau value
\begin{align} \label{Eq:momentum_broadening}
\langle (\Delta \hat{p})^2 \rangle\!(t \gg \ell m/p_0 ) = \frac{\hbar^2}{4 \sigma^2} + \frac{2 m^2 C_0}{p_0^2} .
\end{align}
While both these disorder effects scale inversely with powers of the momentum $p_0$, such that the momentum shift (\ref{Eq:momentum_shift}) is usually negligible, the momentum broadening (\ref{Eq:momentum_broadening}) can, as we show next, even if tiny, potentially still have a relevant impact, mediated by dispersion.

\section{Purity evolution}

To assess the disorder-induced dephasing undergone by a propagating wave packet, we evaluate the purity loss of the disorder-averaged quantum state (see also \cite{Gneiting2016incoherent}). If we evaluate the purity $r(t) \equiv \Tr [\rho(t)^2] = \frac{1}{2 \pi \hbar} \int \dd s \, \dd q \, \chi_t(s,q) \chi_t(-s,-q)$ for solution (\ref{Eq:phase-space_solution}), a Gaussian initial state and Gaussian correlations (\ref{Eq:Gaussian_momentum_transfer_distribution}), and assuming $p_0 \gg \hbar/\ell$, we can approximate, after decorrelation (note that, in the case of purity, decorrelation is determined by both the correlation length $\ell$ and the wave-packet width $\sigma$, i.e., $p_0 t/m \gg \ell,\sigma$),
\begin{align} \label{Eq:purity_loss}
r(t) = 1 - \frac{2 m^2 C_0 \ell}{\hbar^2 p_0^2} \left( \sqrt{\ell^2 + 3 \sigma^2 + \sigma(t)^2} - \ell \right) ,
\end{align}
where $\sigma^2(t) = \sigma^2 + \left(\frac{\hbar t}{2 m \sigma}\right)^2$ describes the dispersive spreading of a free Gaussian wave packet. Remarkably, however, it stems from the disorder influence in Eq.~(\ref{Eq:disorder_influence}). Note that a possible backscattering contribution is neglected in this approximation; moreover, it assumes small purity losses.

Inspecting Eq.~(\ref{Eq:purity_loss}), we find that, in the course of a decorrelation period, purity suffers a loss, which scales with the disorder-induced momentum broadening, cf. Eq.~(\ref{Eq:momentum_broadening}). However, after assuming an intermediate plateau value, a subsequent purity decay sets in (which persists until the approximation breaks down), indicating an increasing susceptibility to dephasing as the wave packet spreads dispersively. We characterize the onset of the dispersion dominance by $\sigma^2(t_{\rm dd}) = 2 \sigma^2$, i.e., $t_{\rm dd} = 2 m \sigma^2/\hbar$. We thus find that, in order to avoid this additional, unbound dephasing, the wave packet must not enter the dispersion-dominated regime, i.e~the duration $t_f$ of the transport task must satisfy $t_f \approx m L/p_0 \lesssim t_{\rm dd} = 2 m \sigma^2/\hbar$, or $\lambdabar_{\rm dB} L \lesssim 2 \sigma^2$, with $L$ the length of the waveguide. We identify this requirement as another benchmark condition for reliable transmittance. Note, however, that, while large wave-packet widths $\sigma$ are favorable from this perspective, there is a tradeoff with the purity loss due to decorrelation. In particular, in the plane-wave limit, the latter dominates.

For example, we now evaluate the functioning of Mach-Zehnder interferometers. The latter have, for instance, been proposed for motional Bell tests with ultracold lithium atoms \cite{Gneiting2008bell, Gneiting2010entangling}. The output probabilities in the final arms (denoted as $\pm$) are given by ${\rm prob}_{\pm}(\varphi) = \frac{1}{2} (1 \pm {\rm Im}[\langle \psi \ket{\psi'} \ee^{\ii \varphi}])$, where $\ket{\psi}$ and $\ket{\psi'}$ denote the wave packets entering the final beam splitter from the long and the short arm, respectively, with $\ket{\psi}$ acquiring an additional phase shift $\varphi$. Optimal functioning, i.e., maximal contrast of the interferometric modulation with varying $\varphi$, presupposes identical input states, $\ket{\psi} = \ket{\psi'}$. Due to disorder-perturbed propagation, however, they differ in general. Performing the disorder average, we obtain $\overline{{\rm prob}}_{\pm}(\varphi) = \frac{1}{2} (1 \pm \frac{r+1}{2} \sin \varphi)$, i.e., the disorder-induced visibility reduction, indicating detrimental leakage between the arms, is quantified by the purity loss of the ensemble-averaged state.
	
Following \cite{Gneiting2008bell, Gneiting2010entangling}, let us assume that a lithium atom propagates at $v_0 \approx 1 \, {\rm cm/s}$, with $\lambdabar_{\rm dB} \approx 1 \, \mu{\rm m}$. To exclude backscattering, the disorder potential (e.g.,~due to fluctuations in the guiding field) should then satisify $\ell \gg 1 \, \mu{\rm m}$, say $\ell = 100 \, \mu{\rm m}$. Moreover, the disorder amplitude must remain well below the kinetic energy, say, $4 m^2 C_0 p_0^{-4} \approx 10^{-5}$. To limit dispersion, we further assume $\sigma \approx \sqrt{\lambdabar_{\rm dB} L/2} \approx 70 \, \mu{\rm m}$. This then yields, after decorrelation, a purity loss of $4 \%$. This may still be tolerable to obtain the required fringe contrast in the interferometers for a successful Bell test.

\section{Numerical test}

We test our theory by comparing it to the numerically exact evolution of the ensemble-averaged state, averaged over $K=250$ disorder realizations. To this end, we propagate (initially) Gaussian wave packets in a 1D Anderson-like model (with lattice spacing $a$ and hopping constant $J$), with Gaussian correlations (\ref{Eq:Gaussian_momentum_transfer_distribution}) ($C_0 = W^2/12$), and in the vicinity of the lower band edge. Our system size is $M = 100$ sites, with periodic boundary conditions. Note that, for larger velocities, at the limits of the quadratic-dispersion approximation, we work with velocity-adapted masses.

\begin{figure}[htb]
	(a) \phantom{aaaaaaaaaaaaaaaaaaaaaaaaaaaaaaaaaaaaaaaaaaaa} \\
	\includegraphics[width=0.7\columnwidth]{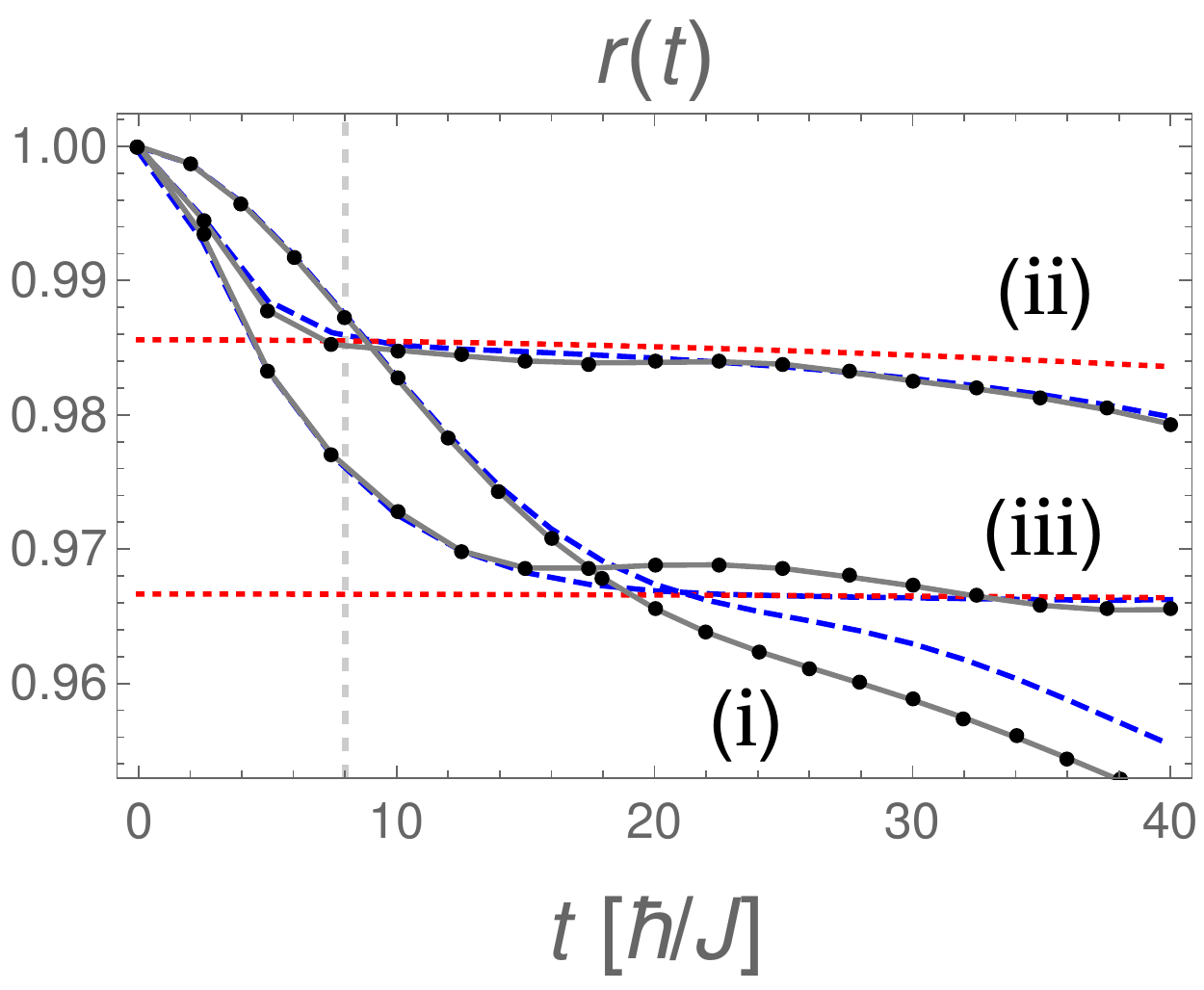} \\
	(b) \phantom{aaaaaaaaaaaaaaaaaaaaaaaaaaaaaaaaaaaaaaaaaaaa} \\
	\includegraphics[width=0.7\columnwidth]{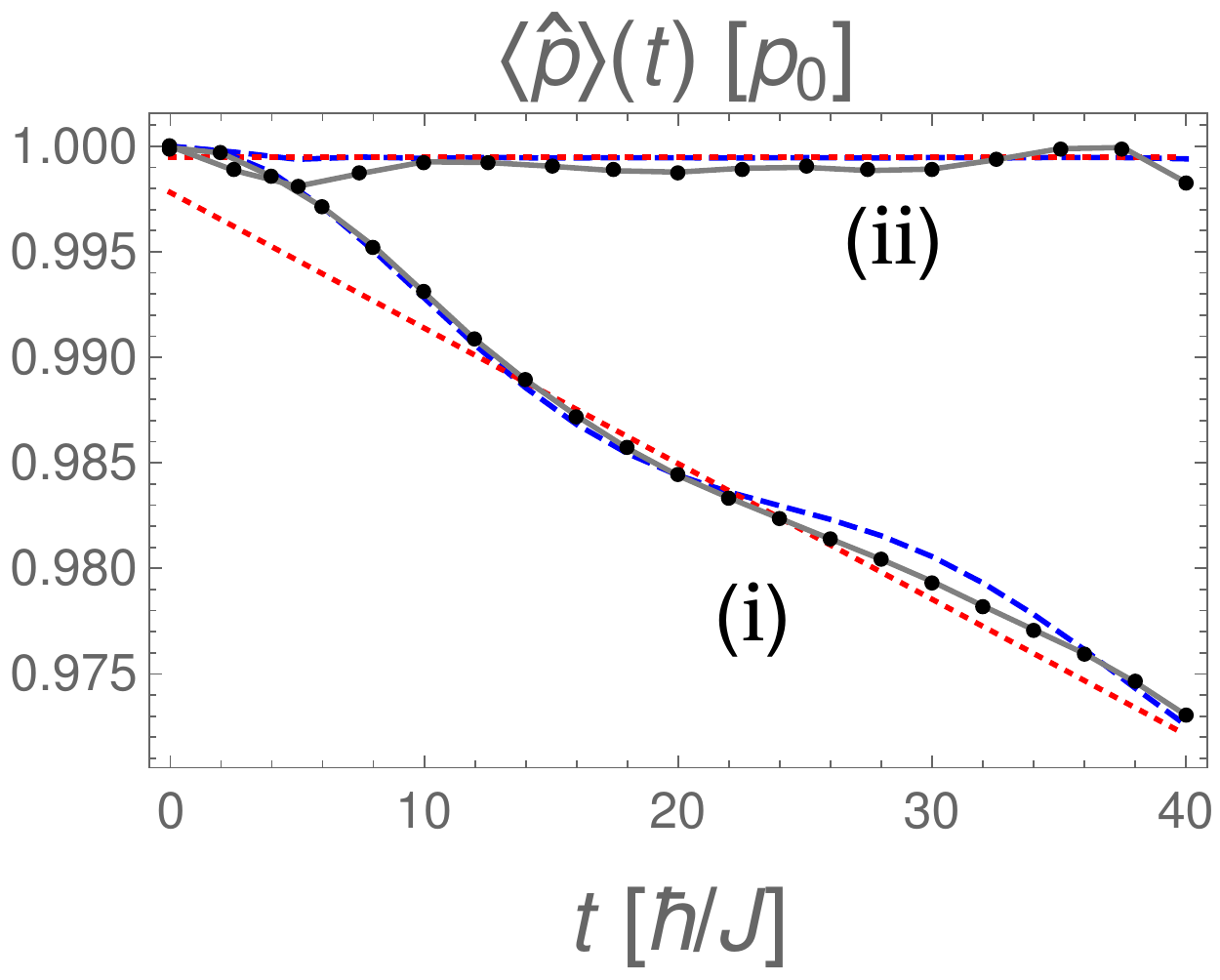}
	\caption{\label{Fig:Numerical_test} (Color online) Propagating wave packet exposed to a weak disorder potential with Gaussian correlations. Shown are the time evolution of the (a) purity $r(t)$ and of the (b) mean momentum $\langle \hat{p} \rangle(t)$ for (i) strong backscattering, (ii) weak backscattering and high dispersion, and (iii) weak backscattering and low dispersion. We compare the numerically exact evolution of the state, averaged over $K=250$ disorder realizations (black dots), the prediction of the propagation equation~(\ref{Eq:phase-space_solution}) (blue dashed curves), and the approximations (\ref{Eq:backscattering_momentum_distribution}) and (\ref{Eq:purity_loss}), respectively (red dotted lines). In all three cases we observe an initial purity loss due to decorrelation. The vertical, grey dashed line indicates the termination of decorrelation for case (ii), i.e., when the purity has reached its intermediate plateau described by (\ref{Eq:purity_loss}). While in case (i) the subsequent purity decay follows from backscattering, in case (ii) it stems from dispersion. This is also reflected in (b), displaying a persistent momentum decrease in (i), while remaining stable in (ii) and (iii). Case iii) is most robust against dephasing on the long run.}
\end{figure}

We compare three benchmark situations: (i) strong backscattering, $\lambdabar_{\rm dB} = \ell$ with $t_{\rm bd} \approx 3.4 \frac{\hbar}{J}$ ($W = 0.05 J$, $\sigma = 10 a$, $\ell = 2 a$), (ii) weak backscattering and high dispersion, with $\lambdabar_{\rm dB} = \ell/4$, $t_{\rm bd} \approx 8.5 \times 10^{5} \frac{\hbar}{J}$, and $t_{\rm dd} \approx 42 \frac{\hbar}{J}$ ($W = 0.1 J$, $\sigma = 5 a$, $\ell = 3 a$), and (iii) weak backscattering and low dispersion, with $\lambdabar_{\rm dB} = \ell/4$, $t_{\rm bd} \approx 8.5 \times 10^{5} \frac{\hbar}{J}$, and $t_{\rm dd} \approx 168 \frac{\hbar}{J}$ ($W = 0.1 J$, $\sigma = 10 a$, $\ell = 3 a$). In Fig.~\ref{Fig:Numerical_test} we compare the numerically exact evolution of the mean momentum and the purity with the prediction of the propagation equation~(\ref{Eq:phase-space_solution}) and the approximations (\ref{Eq:backscattering_momentum_distribution}) and (\ref{Eq:purity_loss}), respectively.

We find that Eq.~(\ref{Eq:phase-space_solution}) correctly reproduces the disorder-induced dynamics, when taking into account statistical deviations due to the moderate number ($\sim$ 250) of realizations. In all three cases we observe an initial purity loss due to decorrelation. While in case (i) the subsequent purity decay follows from backscattering, in (ii) it stems from dispersion. This is confirmed by the evolution of the mean momentum, displaying a persistent decrease in case (i), while remaining stable in (ii) and (iii). In case (iii) the subsequent purity decay is weakest. Thus, irrespective of the larger purity loss [compared to (ii)] in the decorrelation period, on the long run case (iii) is better for reliable quantum state transmittance. Let us remind the reader that the validity of Eq.~(\ref{Eq:phase-space_solution}), which is perturbative in $C_0/\left( p_0^2/2 m \right)^2$, ceases when two or more scattering events become relevant, i.e., beyond the scope of reliable transport.

Note that Eq.~(\ref{Eq:purity_loss}) underestimates the dispersion-induced purity lost in case (ii). This is due to the neglected higher-order corrections in $\hbar/(p_0 \ell)$, which remain significant at $\lambdabar_{\rm dB} = \ell/4$ and the considered times close to $t_{\rm dd}$. We further remark that the momentum shift (\ref{Eq:momentum_shift}), which is $\sim 0.1 \%$, is overlaid by statistical fluctuations. The broadening of the momentum variance, Eq.~(\ref{Eq:momentum_broadening}), however, is reproduced well (not shown), yielding an increase of about (i) $20 \%$, (ii) $6 \%$, and (iii) $26 \%$.

\section{Coupled disorder channels}

Hereafter, we introduce the coupled disorder channels method, which can be used to derive the propagation equation~(\ref{Eq:phase-space_solution}). In analogy to the coupled channels equations used in atomic theory to describe, e.g., the dynamical influence of closed channels on an open channel \cite{Stoof1988spin, Koehler2006production, Gneiting2010molecular}, we derive a set of dynamical equations describing the impact of disorder realizations on the evolution of the ensemble-averaged state. The following holds for arbitrary disorder configurations.

In order to obtain the desired decomposition of the dynamics into distinct disorder channels, we represent the state $\rho_{\varepsilon}$ of each specific disorder realization in terms of the ensemble-averaged state $\overline{\rho}$ and its individual offset $\Delta\rho_{\varepsilon}$ from the latter: $\rho_{\varepsilon} = \overline{\rho} + \Delta\rho_{\varepsilon}$. If, in addition, we define $\hat{H}_{\varepsilon} = \hat{\overline{H}} + \hat{V}_{\varepsilon}$, where $\hat{\overline{H}} \equiv \int \dd \varepsilon \, p_{\varepsilon} \, \hat{H}_{\varepsilon}$ and $\int \dd \varepsilon \, p_{\varepsilon} \, \hat{V}_{\varepsilon} = 0$, we obtain, with the von Neumann equation, coupled evolution equations for the ensemble-averaged state $\overline{\rho}$ and the offsets $\Delta\rho_{\varepsilon}$:
\begin{subequations} \label{Eq:coupled_disorder_channels}
\begin{align}
\partial_t \overline{\rho}(t) =& -\frac{\ii}{\hbar} [\hat{\overline{H}}, \overline{\rho}(t)] - \frac{\ii}{\hbar} \int \dd \varepsilon \, p_{\varepsilon} \, [\hat{V}_{\varepsilon}, \Delta\rho_{\varepsilon}(t)] , \label{Eq:average_evolution} \\
\partial_t \Delta\rho_{\varepsilon}(t) =& -\frac{\ii}{\hbar} [\hat{H}_{\varepsilon}, \Delta\rho_{\varepsilon}(t)] -\frac{\ii}{\hbar} [\hat{V}_{\varepsilon},\overline{\rho}(t)] \nonumber \label{Eq:offset_evolution} \\
 &+\frac{\ii}{\hbar} \int \dd \varepsilon' \, p_{\varepsilon'} \, [\hat{V}_{\varepsilon'}, \Delta\rho_{\varepsilon'}(t)] .
\end{align}
\end{subequations}
The corresponding initial conditions are $\overline{\rho}(t=0) = \rho_0$ and $\Delta\rho_{\varepsilon}(t=0) = 0, \, \forall \varepsilon$. Note that (\ref{Eq:offset_evolution}) encompasses a coupling between different disorder realizations, mediated by their collective influence on the ensemble-averaged state $\overline{\rho}(t)$. Moreover, in contrast to the ensemble average $\overline{\rho}(t)$, the offsets $\Delta\rho_{\varepsilon}(t)$ do not describe normalized quantum states.

The coupled disorder channel equations (\ref{Eq:coupled_disorder_channels}) are exact and capture the full dynamics of the disorder ensemble, including the time evolution of single disorder realizations. More importantly, however, they allow us to develop systematic approximation methods to obtain closed evolution equations for the ensemble-averaged state $\overline{\rho}(t)$.

Note that the coupled disorder channel equations (\ref{Eq:coupled_disorder_channels}) can be related to the Nakajima-Zwanzig projection operator technique \cite{Nakajima1959quantum, Zwanzig1960ensemble, Breuer2002theory}, generalizing the latter to the case of many irrelevant components.

Solving Eqs.~(\ref{Eq:offset_evolution}) formally in the Green's formalism yields recursive, time-nonlocal integral representations for the offsets $\Delta\rho_{\varepsilon}(t)$. In the perturbative limit, one derives, with Eq.~(\ref{Eq:average_evolution}), a closed, time-local quantum master equation in Lindblad form for $\overline{\rho}(t)$, which is second order in $\hat{V}_{\varepsilon}$:
\begin{subequations} \label{Eq:perturbative_master_equation}
\begin{align}
\partial_t \overline{\rho}(t) =& -\frac{\ii}{\hbar} [\hat{H}_{\rm eff}(t), \overline{\rho}(t)] \nonumber \\
 & +\sum_{\alpha \in \{ \pm 1 \}} \frac{2 \alpha}{\hbar^2} \int \dd \varepsilon \, p_{\varepsilon} \int_{0}^{t} \dd t' \mathcal{L}\big(\hat{L}_{\varepsilon, t'}^{(\alpha)}, \overline{\rho}(t)\big) .
\end{align}
Here, $\mathcal{L}(\hat{L}, \rho) \equiv \hat{L} \rho \hat{L}^{\dagger} - \frac{1}{2} \hat{L}^{\dagger} \hat{L} \rho - \frac{1}{2} \rho \hat{L}^{\dagger} \hat{L}$. The (in general time-dependent) effective Hamiltonian $H_{\rm eff}(t) = H_{\rm eff}^{\dagger}(t)$ and Lindblad operators $\hat{L}_{\varepsilon, t}^{(\alpha)}$ are given by
	\begin{align}
	\hat{H}_{\rm eff}(t) &= \hat{\overline{H}} -\frac{\ii}{2 \hbar} \int \dd \varepsilon \, p_{\varepsilon} \int_{0}^{t} \dd t' \, [\hat{V}_{\varepsilon}, \hat{\tilde{V}}_{\varepsilon}(t')] , \nonumber \\
	\hat{L}_{\varepsilon, t}^{(\alpha)} &= \frac{1}{2} \big[\hat{V}_{\varepsilon} + \alpha \, \hat{\tilde{V}}_{\varepsilon}(t) \big] ,
	\end{align}
\end{subequations}
where $\hat{\tilde{V}}_{\varepsilon}(t) = \hat{\overline{U}}(t) \hat{V}_{\varepsilon} \hat{\overline{U}}(t)^{\dagger}$ and $\hat{\overline{U}}(t) = \exp(- \ii \hat{\overline{H}} t/\hbar)$. Note how Eq.~(\ref{Eq:perturbative_master_equation}) consistently separates coherent and incoherent contributions to the evolution. Moreover, we remark that the $\alpha = -1$ term in Eq.~(\ref{Eq:perturbative_master_equation}) refers to the feedback of coherence into the system, while the corresponding incoherent process, $\hat{L}_{\varepsilon, t}^{(-)}$, only builds up slowly in time, $\hat{L}_{\varepsilon, t=0}^{(-)} = 0$, in agreement with the positivity of the evolution. Let us stress that taking the limit $t \rightarrow \infty$ in the integral, corresponding to a Markov approximation, would in general not only neglect essential disorder effects, but also lead to wrong results.

Solving the quantum master equation~(\ref{Eq:perturbative_master_equation}), following similar steps as in \cite{Gneiting2016incoherent}, for a disorder-perturbed, but otherwise free, material particle, yields the propagation equation~(\ref{Eq:phase-space_solution}). To this end, it is helpful to  (intermediately) switch into a comoving frame, described by the replacement $\hat{p}^2/2 m \rightarrow \hat{p}^2/2 m + p_0 \hat{p}/m$ in the averaged Hamiltonian. Solution~(\ref{Eq:phase-space_solution}) results exactly from Eq.~(8) and thus is manifestly superior to a standard Born approximation, obtained by expanding Eq.~(\ref{Eq:phase-space_solution}) to first order in $C_0$.

\section{Conclusions}

We analyzed the transport of massive quantum particles through disorder-perturbed single-mode waveguides. To this end, we presented the propagation equation (\ref{Eq:phase-space_solution}), which temporally resolves the impact of weak disorder on the full ensemble-averaged quantum state. It is valid for kinetic energies sufficiently large compared to the disorder strength and up to the second backscattering event. Propagation equation (\ref{Eq:phase-space_solution}) describes, within its validity range, the disorder effects, on all diagonal and off-diagonal matrix elements, or, equivalently, on state moments of any order, position and momentum, for arbitrary input states with the required properties and sufficiently well-behaved disorder correlations. In that sense, while reproducing known disorder effects, it provides a comprehensive description of the propagation in the presence of disorder, as it may be required in order to assess the impact of disorder when particles function as carriers of quantum information.

For examples, we elaborated the first two momentum moments for Gaussian input states and Gaussian disorder correlations, identifying, in the course of a decorrelation period, a momentum reduction and a momentum broadening. This goes along with a purity loss, which, in the dispersion-dominated regime, persists even in the absence of backscattering. Consequently, we identify two benchmark requirements for reliable quantum state transmittance: $p_0 \ell/\hbar \gg 1$ or $\lambdabar_{\rm dB} \ll \ell$ (weak backscattering) and $t_f \approx m L/p_0 \lesssim 2 m \sigma^2/\hbar$ or $\lambdabar_{\rm dB} L \lesssim 2 \sigma^2$ (low dispersion).

We expect that our description, which accurately captures any disorder effect, classical or quantum, in the regime of reliable transport, may generally be relevant for devices or experiments which rely on the precise transport of quantum states, i.e., where details of state propagation are relevant and a comprehensive state description is imperative. In that sense, it may complement other established methods to treat disordered quantum systems, e.g., based on Green's functions. Attractive systems for scrutiny tests are ultracold atoms in optical waveguides \cite{Billy2008direct, Mueller2015suppression}, microwave waveguides \cite{Kuhl2000experimental, Fernandez2014beyond}, and classical light in the paraxial approximation \cite{Schwartz2007transport, Segev2013anderson}, which all allow high control over disorder properties and state readout with excellent spatial resolution.

While the perturbative master equation (\ref{Eq:perturbative_master_equation}) can also be applied to a variety of other cases, such as transport in spin baths or mobility edges in higher dimensions, the method of coupled disorder channels (\ref{Eq:coupled_disorder_channels}) may also be used to systematically derive quantum master equations beyond the perturbative limit considered here, then possibly encompassing higher-order disorder effects such as coherent backscattering.

\paragraph*{Acknowledgments.}

This research was partially supported by the RIKEN iTHES Project, the MURI Center for Dynamic Magneto-Optics via the AFOSR award number FA9550-14-1-0040, the Japan Society for the Promotion of Science (KAKENHI), the IMPACT program of JST, JSPS-RFBR grant No.17-52-50023, CREST grant No. JPMJCR1676, a Grant-in-Aid for Scientific Research (A), and a grant from the John Templeton Foundation.

\bibliography{literature}

\begin{thebibliography}{34}%
\makeatletter
\providecommand \@ifxundefined [1]{%
 \@ifx{#1\undefined}
}%
\providecommand \@ifnum [1]{%
 \ifnum #1\expandafter \@firstoftwo
 \else \expandafter \@secondoftwo
 \fi
}%
\providecommand \@ifx [1]{%
 \ifx #1\expandafter \@firstoftwo
 \else \expandafter \@secondoftwo
 \fi
}%
\providecommand \natexlab [1]{#1}%
\providecommand \enquote  [1]{``#1''}%
\providecommand \bibnamefont  [1]{#1}%
\providecommand \bibfnamefont [1]{#1}%
\providecommand \citenamefont [1]{#1}%
\providecommand \href@noop [0]{\@secondoftwo}%
\providecommand \href [0]{\begingroup \@sanitize@url \@href}%
\providecommand \@href[1]{\@@startlink{#1}\@@href}%
\providecommand \@@href[1]{\endgroup#1\@@endlink}%
\providecommand \@sanitize@url [0]{\catcode `\\12\catcode `\$12\catcode
  `\&12\catcode `\#12\catcode `\^12\catcode `\_12\catcode `\%12\relax}%
\providecommand \@@startlink[1]{}%
\providecommand \@@endlink[0]{}%
\providecommand \url  [0]{\begingroup\@sanitize@url \@url }%
\providecommand \@url [1]{\endgroup\@href {#1}{\urlprefix }}%
\providecommand \urlprefix  [0]{URL }%
\providecommand \Eprint [0]{\href }%
\providecommand \doibase [0]{http://dx.doi.org/}%
\providecommand \selectlanguage [0]{\@gobble}%
\providecommand \bibinfo  [0]{\@secondoftwo}%
\providecommand \bibfield  [0]{\@secondoftwo}%
\providecommand \translation [1]{[#1]}%
\providecommand \BibitemOpen [0]{}%
\providecommand \bibitemStop [0]{}%
\providecommand \bibitemNoStop [0]{.\EOS\space}%
\providecommand \EOS [0]{\spacefactor3000\relax}%
\providecommand \BibitemShut  [1]{\csname bibitem#1\endcsname}%
\let\auto@bib@innerbib\@empty
\bibitem [{\citenamefont {Lifshits}\ \emph {et~al.}(1988)\citenamefont
  {Lifshits}, \citenamefont {Gredeskul},\ and\ \citenamefont
  {Pastur}}]{Lifshits1988introduction}%
  \BibitemOpen
  \bibfield  {author} {\bibinfo {author} {\bibfnamefont {I.~M.}\ \bibnamefont
  {Lifshits}}, \bibinfo {author} {\bibfnamefont {S.~A.}\ \bibnamefont
  {Gredeskul}}, \ and\ \bibinfo {author} {\bibfnamefont {L.~A.}\ \bibnamefont
  {Pastur}},\ }\href@noop {} {\emph {\bibinfo {title} {Introduction to the
  Theory of Disordered Systems}}}\ (\bibinfo  {publisher}
  {Wiley-Interscience},\ \bibinfo {year} {1988})\BibitemShut {NoStop}%
\bibitem [{\citenamefont {Rammer}(1991)}]{Rammer1991quantum}%
  \BibitemOpen
  \bibfield  {author} {\bibinfo {author} {\bibfnamefont {J.}~\bibnamefont
  {Rammer}},\ }\bibfield  {title} {\enquote {\bibinfo {title} {Quantum
  transport theory of electrons in solids: A single-particle approach},}\
  }\href@noop {} {\bibfield  {journal} {\bibinfo  {journal} {Rev. Mod. Phys.}\
  }\textbf {\bibinfo {volume} {63}},\ \bibinfo {pages} {781--817} (\bibinfo
  {year} {1991})}\BibitemShut {NoStop}%
\bibitem [{\citenamefont {Beenakker}(1997)}]{Beenakker1997random}%
  \BibitemOpen
  \bibfield  {author} {\bibinfo {author} {\bibfnamefont {C.~W.~J.}\
  \bibnamefont {Beenakker}},\ }\bibfield  {title} {\enquote {\bibinfo {title}
  {Random-matrix theory of quantum transport},}\ }\href {\doibase
  10.1103/RevModPhys.69.731} {\bibfield  {journal} {\bibinfo  {journal} {Rev.
  Mod. Phys.}\ }\textbf {\bibinfo {volume} {69}},\ \bibinfo {pages} {731--808}
  (\bibinfo {year} {1997})}\BibitemShut {NoStop}%
\bibitem [{\citenamefont {Akkermans}\ and\ \citenamefont
  {Montambaux}(2007)}]{Akkermans2007mesoscopic}%
  \BibitemOpen
  \bibfield  {author} {\bibinfo {author} {\bibfnamefont {E.}~\bibnamefont
  {Akkermans}}\ and\ \bibinfo {author} {\bibfnamefont {G.}~\bibnamefont
  {Montambaux}},\ }\href@noop {} {\emph {\bibinfo {title} {Mesoscopic Physics
  of Electrons and Photons}}}\ (\bibinfo  {publisher} {Cambridge Univ. Press},\
  \bibinfo {year} {2007})\BibitemShut {NoStop}%
\bibitem [{\citenamefont {Nishiguchi}\ \emph {et~al.}(1993)\citenamefont
  {Nishiguchi}, \citenamefont {Tamura},\ and\ \citenamefont
  {Nori}}]{Nishiguchi1993phonon}%
  \BibitemOpen
  \bibfield  {author} {\bibinfo {author} {\bibfnamefont {N.}~\bibnamefont
  {Nishiguchi}}, \bibinfo {author} {\bibfnamefont {S.-I.}\ \bibnamefont
  {Tamura}}, \ and\ \bibinfo {author} {\bibfnamefont {F.}~\bibnamefont
  {Nori}},\ }\bibfield  {title} {\enquote {\bibinfo {title}
  {Phonon-transmission rate, fluctuations, and localization in random
  semiconductor superlattices: {G}reen's-function approach},}\ }\href {\doibase
  10.1103/PhysRevB.48.2515} {\bibfield  {journal} {\bibinfo  {journal} {Phys.
  Rev. B}\ }\textbf {\bibinfo {volume} {48}},\ \bibinfo {pages} {2515--2528}
  (\bibinfo {year} {1993})}\BibitemShut {NoStop}%
\bibitem [{\citenamefont {Lin}\ and\ \citenamefont
  {Nori}(1996)}]{Lin1996analytical}%
  \BibitemOpen
  \bibfield  {author} {\bibinfo {author} {\bibfnamefont {Y.-L.}\ \bibnamefont
  {Lin}}\ and\ \bibinfo {author} {\bibfnamefont {F.}~\bibnamefont {Nori}},\
  }\bibfield  {title} {\enquote {\bibinfo {title} {Analytical results on
  quantum interference and magnetoconductance for strongly localized electrons
  in a magnetic field: Exact summation of forward-scattering paths},}\ }\href
  {\doibase 10.1103/PhysRevB.53.15543} {\bibfield  {journal} {\bibinfo
  {journal} {Phys. Rev. B}\ }\textbf {\bibinfo {volume} {53}},\ \bibinfo
  {pages} {15543--15562} (\bibinfo {year} {1996})}\BibitemShut {NoStop}%
\bibitem [{\citenamefont {Friesen}\ \emph {et~al.}(2007)\citenamefont
  {Friesen}, \citenamefont {Biswas}, \citenamefont {Hu},\ and\ \citenamefont
  {Lidar}}]{Friesen2007efficient}%
  \BibitemOpen
  \bibfield  {author} {\bibinfo {author} {\bibfnamefont {M.}~\bibnamefont
  {Friesen}}, \bibinfo {author} {\bibfnamefont {A.}~\bibnamefont {Biswas}},
  \bibinfo {author} {\bibfnamefont {X.}~\bibnamefont {Hu}}, \ and\ \bibinfo
  {author} {\bibfnamefont {D.}~\bibnamefont {Lidar}},\ }\bibfield  {title}
  {\enquote {\bibinfo {title} {Efficient multiqubit entanglement via a spin
  bus},}\ }\href {\doibase 10.1103/PhysRevLett.98.230503} {\bibfield  {journal}
  {\bibinfo  {journal} {Phys. Rev. Lett.}\ }\textbf {\bibinfo {volume} {98}},\
  \bibinfo {pages} {230503} (\bibinfo {year} {2007})}\BibitemShut {NoStop}%
\bibitem [{\citenamefont {Cronin}\ \emph {et~al.}(2009)\citenamefont {Cronin},
  \citenamefont {Schmiedmayer},\ and\ \citenamefont
  {Pritchard}}]{Cronin2009optics}%
  \BibitemOpen
  \bibfield  {author} {\bibinfo {author} {\bibfnamefont {A.~D.}\ \bibnamefont
  {Cronin}}, \bibinfo {author} {\bibfnamefont {J.}~\bibnamefont
  {Schmiedmayer}}, \ and\ \bibinfo {author} {\bibfnamefont {D.~E.}\
  \bibnamefont {Pritchard}},\ }\bibfield  {title} {\enquote {\bibinfo {title}
  {Optics and interferometry with atoms and molecules},}\ }\href {\doibase
  10.1103/RevModPhys.81.1051} {\bibfield  {journal} {\bibinfo  {journal} {Rev.
  Mod. Phys.}\ }\textbf {\bibinfo {volume} {81}},\ \bibinfo {pages}
  {1051--1129} (\bibinfo {year} {2009})}\BibitemShut {NoStop}%
\bibitem [{\citenamefont {Buluta}\ \emph {et~al.}(2011)\citenamefont {Buluta},
  \citenamefont {Ashhab},\ and\ \citenamefont {Nori}}]{Buluta2011natural}%
  \BibitemOpen
  \bibfield  {author} {\bibinfo {author} {\bibfnamefont {I.}~\bibnamefont
  {Buluta}}, \bibinfo {author} {\bibfnamefont {S.}~\bibnamefont {Ashhab}}, \
  and\ \bibinfo {author} {\bibfnamefont {F.}~\bibnamefont {Nori}},\ }\bibfield
  {title} {\enquote {\bibinfo {title} {Natural and artificial atoms for quantum
  computation},}\ }\href {http://stacks.iop.org/0034-4885/74/i=10/a=104401}
  {\bibfield  {journal} {\bibinfo  {journal} {Rep.~Progr.~Phys.}\ }\textbf
  {\bibinfo {volume} {74}},\ \bibinfo {pages} {104401} (\bibinfo {year}
  {2011})}\BibitemShut {NoStop}%
\bibitem [{\citenamefont {Huang}\ and\ \citenamefont
  {Hu}(2013)}]{Huang2013spin}%
  \BibitemOpen
  \bibfield  {author} {\bibinfo {author} {\bibfnamefont {P.}~\bibnamefont
  {Huang}}\ and\ \bibinfo {author} {\bibfnamefont {X.}~\bibnamefont {Hu}},\
  }\bibfield  {title} {\enquote {\bibinfo {title} {Spin qubit relaxation in a
  moving quantum dot},}\ }\href {\doibase 10.1103/PhysRevB.88.075301}
  {\bibfield  {journal} {\bibinfo  {journal} {Phys. Rev. B}\ }\textbf {\bibinfo
  {volume} {88}},\ \bibinfo {pages} {075301} (\bibinfo {year}
  {2013})}\BibitemShut {NoStop}%
\bibitem [{\citenamefont {Zhao}\ \emph {et~al.}(2016)\citenamefont {Zhao},
  \citenamefont {Huang},\ and\ \citenamefont {Hu}}]{Zhao2016doppler}%
  \BibitemOpen
  \bibfield  {author} {\bibinfo {author} {\bibfnamefont {X.}~\bibnamefont
  {Zhao}}, \bibinfo {author} {\bibfnamefont {P.}~\bibnamefont {Huang}}, \ and\
  \bibinfo {author} {\bibfnamefont {X.}~\bibnamefont {Hu}},\ }\bibfield
  {title} {\enquote {\bibinfo {title} {Doppler effect induced spin relaxation
  boom},}\ }\href@noop {} {\bibfield  {journal} {\bibinfo  {journal}
  {Scientific Reports}\ }\textbf {\bibinfo {volume} {6}},\ \bibinfo {pages}
  {23169} (\bibinfo {year} {2016})}\BibitemShut {NoStop}%
\bibitem [{\citenamefont {Lekitsch}\ \emph {et~al.}(2017)\citenamefont
  {Lekitsch}, \citenamefont {Weidt}, \citenamefont {Fowler}, \citenamefont
  {M{\o}lmer}, \citenamefont {Devitt}, \citenamefont {Wunderlich},\ and\
  \citenamefont {Hensinger}}]{Lekitsche2017blueprint}%
  \BibitemOpen
  \bibfield  {author} {\bibinfo {author} {\bibfnamefont {B.}~\bibnamefont
  {Lekitsch}}, \bibinfo {author} {\bibfnamefont {S.}~\bibnamefont {Weidt}},
  \bibinfo {author} {\bibfnamefont {A.~G.}\ \bibnamefont {Fowler}}, \bibinfo
  {author} {\bibfnamefont {K.}~\bibnamefont {M{\o}lmer}}, \bibinfo {author}
  {\bibfnamefont {S.~J.}\ \bibnamefont {Devitt}}, \bibinfo {author}
  {\bibfnamefont {C.}~\bibnamefont {Wunderlich}}, \ and\ \bibinfo {author}
  {\bibfnamefont {W.~K.}\ \bibnamefont {Hensinger}},\ }\bibfield  {title}
  {\enquote {\bibinfo {title} {Blueprint for a microwave trapped ion quantum
  computer},}\ }\href@noop {} {\bibfield  {journal} {\bibinfo  {journal}
  {Science Advances}\ }\textbf {\bibinfo {volume} {3}},\ \bibinfo {pages}
  {e1601540} (\bibinfo {year} {2017})}\BibitemShut {NoStop}%
\bibitem [{\citenamefont {Mair}\ \emph {et~al.}(2001)\citenamefont {Mair},
  \citenamefont {Vaziri}, \citenamefont {Weihs},\ and\ \citenamefont
  {Zeilinger}}]{Mair2001entanglement}%
  \BibitemOpen
  \bibfield  {author} {\bibinfo {author} {\bibfnamefont {A.}~\bibnamefont
  {Mair}}, \bibinfo {author} {\bibfnamefont {A.}~\bibnamefont {Vaziri}},
  \bibinfo {author} {\bibfnamefont {G.}~\bibnamefont {Weihs}}, \ and\ \bibinfo
  {author} {\bibfnamefont {A.}~\bibnamefont {Zeilinger}},\ }\bibfield  {title}
  {\enquote {\bibinfo {title} {Entanglement of the orbital angular momentum
  states of photons},}\ }\href@noop {} {\bibfield  {journal} {\bibinfo
  {journal} {Nature}\ }\textbf {\bibinfo {volume} {412}},\ \bibinfo {pages}
  {313--316} (\bibinfo {year} {2001})}\BibitemShut {NoStop}%
\bibitem [{\citenamefont {Brezger}\ \emph {et~al.}(2002)\citenamefont
  {Brezger}, \citenamefont {Hackerm\"uller}, \citenamefont {Uttenthaler},
  \citenamefont {Petschinka}, \citenamefont {Arndt},\ and\ \citenamefont
  {Zeilinger}}]{Brezger2002matter}%
  \BibitemOpen
  \bibfield  {author} {\bibinfo {author} {\bibfnamefont {B.}~\bibnamefont
  {Brezger}}, \bibinfo {author} {\bibfnamefont {L.}~\bibnamefont
  {Hackerm\"uller}}, \bibinfo {author} {\bibfnamefont {S.}~\bibnamefont
  {Uttenthaler}}, \bibinfo {author} {\bibfnamefont {J.}~\bibnamefont
  {Petschinka}}, \bibinfo {author} {\bibfnamefont {M.}~\bibnamefont {Arndt}}, \
  and\ \bibinfo {author} {\bibfnamefont {A.}~\bibnamefont {Zeilinger}},\
  }\bibfield  {title} {\enquote {\bibinfo {title} {Matter-wave interferometer
  for large molecules},}\ }\href {\doibase 10.1103/PhysRevLett.88.100404}
  {\bibfield  {journal} {\bibinfo  {journal} {Phys.~Rev.~Lett.}\ }\textbf
  {\bibinfo {volume} {88}},\ \bibinfo {pages} {100404} (\bibinfo {year}
  {2002})}\BibitemShut {NoStop}%
\bibitem [{\citenamefont {Khakimov}\ \emph {et~al.}(2016)\citenamefont
  {Khakimov}, \citenamefont {Henson}, \citenamefont {Shin}, \citenamefont
  {Hodgman}, \citenamefont {Dall}, \citenamefont {Baldwin},\ and\ \citenamefont
  {Truscott}}]{Khakimov2016ghost}%
  \BibitemOpen
  \bibfield  {author} {\bibinfo {author} {\bibfnamefont {R.~I.}\ \bibnamefont
  {Khakimov}}, \bibinfo {author} {\bibfnamefont {B.~M.}\ \bibnamefont
  {Henson}}, \bibinfo {author} {\bibfnamefont {D.~K.}\ \bibnamefont {Shin}},
  \bibinfo {author} {\bibfnamefont {S.~S.}\ \bibnamefont {Hodgman}}, \bibinfo
  {author} {\bibfnamefont {R.~G.}\ \bibnamefont {Dall}}, \bibinfo {author}
  {\bibfnamefont {K.~G.~H.}\ \bibnamefont {Baldwin}}, \ and\ \bibinfo {author}
  {\bibfnamefont {A.~G.}\ \bibnamefont {Truscott}},\ }\bibfield  {title}
  {\enquote {\bibinfo {title} {Ghost imaging with atoms},}\ }\href@noop {}
  {\bibfield  {journal} {\bibinfo  {journal} {Nature}\ }\textbf {\bibinfo
  {volume} {540}},\ \bibinfo {pages} {100--103} (\bibinfo {year}
  {2016})}\BibitemShut {NoStop}%
\bibitem [{\citenamefont {Gneiting}\ \emph {et~al.}(2016)\citenamefont
  {Gneiting}, \citenamefont {Anger},\ and\ \citenamefont
  {Buchleitner}}]{Gneiting2016incoherent}%
  \BibitemOpen
  \bibfield  {author} {\bibinfo {author} {\bibfnamefont {C.}~\bibnamefont
  {Gneiting}}, \bibinfo {author} {\bibfnamefont {F.~R.}\ \bibnamefont {Anger}},
  \ and\ \bibinfo {author} {\bibfnamefont {A.}~\bibnamefont {Buchleitner}},\
  }\bibfield  {title} {\enquote {\bibinfo {title} {Incoherent ensemble dynamics
  in disordered systems},}\ }\href@noop {} {\bibfield  {journal} {\bibinfo
  {journal} {Phys. Rev. A}\ }\textbf {\bibinfo {volume} {93}},\ \bibinfo
  {pages} {032139} (\bibinfo {year} {2016})}\BibitemShut {NoStop}%
\bibitem [{\citenamefont {Kropf}\ \emph {et~al.}(2016)\citenamefont {Kropf},
  \citenamefont {Gneiting},\ and\ \citenamefont
  {Buchleitner}}]{Kropf2016effective}%
  \BibitemOpen
  \bibfield  {author} {\bibinfo {author} {\bibfnamefont {C.~M.}\ \bibnamefont
  {Kropf}}, \bibinfo {author} {\bibfnamefont {C.}~\bibnamefont {Gneiting}}, \
  and\ \bibinfo {author} {\bibfnamefont {A.}~\bibnamefont {Buchleitner}},\
  }\bibfield  {title} {\enquote {\bibinfo {title} {Effective dynamics of
  disordered quantum systems},}\ }\href@noop {} {\bibfield  {journal} {\bibinfo
   {journal} {Phys. Rev. X}\ }\textbf {\bibinfo {volume} {6}},\ \bibinfo
  {pages} {031023} (\bibinfo {year} {2016})}\BibitemShut {NoStop}%
\bibitem [{\citenamefont {Schleich}(2011)}]{Schleich2011quantum}%
  \BibitemOpen
  \bibfield  {author} {\bibinfo {author} {\bibfnamefont {W.~P.}\ \bibnamefont
  {Schleich}},\ }\href@noop {} {\emph {\bibinfo {title} {Quantum Optics in
  Phase Space}}}\ (\bibinfo  {publisher} {John Wiley \& Sons},\ \bibinfo {year}
  {2011})\BibitemShut {NoStop}%
\bibitem [{\citenamefont {Gneiting}\ \emph {et~al.}(2013)\citenamefont
  {Gneiting}, \citenamefont {Fischer},\ and\ \citenamefont
  {Hornberger}}]{Gneiting2013quantum}%
  \BibitemOpen
  \bibfield  {author} {\bibinfo {author} {\bibfnamefont {C.}~\bibnamefont
  {Gneiting}}, \bibinfo {author} {\bibfnamefont {T.}~\bibnamefont {Fischer}}, \
  and\ \bibinfo {author} {\bibfnamefont {K.}~\bibnamefont {Hornberger}},\
  }\bibfield  {title} {\enquote {\bibinfo {title} {Quantum phase-space
  representation for curved configuration spaces},}\ }\href@noop {} {\bibfield
  {journal} {\bibinfo  {journal} {Phys. Rev. A}\ }\textbf {\bibinfo {volume}
  {88}},\ \bibinfo {pages} {062117} (\bibinfo {year} {2013})}\BibitemShut
  {NoStop}%
\bibitem [{\citenamefont {Izrailev}\ \emph {et~al.}(2012)\citenamefont
  {Izrailev}, \citenamefont {Krokhin},\ and\ \citenamefont
  {Makarov}}]{Izrailev2012anomalous}%
  \BibitemOpen
  \bibfield  {author} {\bibinfo {author} {\bibfnamefont {F.M.}\ \bibnamefont
  {Izrailev}}, \bibinfo {author} {\bibfnamefont {A.A.}\ \bibnamefont
  {Krokhin}}, \ and\ \bibinfo {author} {\bibfnamefont {N.M.}\ \bibnamefont
  {Makarov}},\ }\bibfield  {title} {\enquote {\bibinfo {title} {Anomalous
  localization in low-dimensional systems with correlated disorder},}\ }\href
  {\doibase http://dx.doi.org/10.1016/j.physrep.2011.11.002} {\bibfield
  {journal} {\bibinfo  {journal} {Phys. Rep.}\ }\textbf {\bibinfo {volume}
  {512}},\ \bibinfo {pages} {125 -- 254} (\bibinfo {year} {2012})}\BibitemShut
  {NoStop}%
\bibitem [{\citenamefont {Gneiting}\ and\ \citenamefont
  {Hornberger}(2008)}]{Gneiting2008bell}%
  \BibitemOpen
  \bibfield  {author} {\bibinfo {author} {\bibfnamefont {C.}~\bibnamefont
  {Gneiting}}\ and\ \bibinfo {author} {\bibfnamefont {K.}~\bibnamefont
  {Hornberger}},\ }\bibfield  {title} {\enquote {\bibinfo {title} {Bell test
  for the free motion of material particles},}\ }\href@noop {} {\bibfield
  {journal} {\bibinfo  {journal} {Phys. Rev. Lett.}\ }\textbf {\bibinfo
  {volume} {101}},\ \bibinfo {pages} {260503} (\bibinfo {year}
  {2008})}\BibitemShut {NoStop}%
\bibitem [{\citenamefont {Gneiting}\ and\ \citenamefont
  {Hornberger}(2010{\natexlab{a}})}]{Gneiting2010entangling}%
  \BibitemOpen
  \bibfield  {author} {\bibinfo {author} {\bibfnamefont {C.}~\bibnamefont
  {Gneiting}}\ and\ \bibinfo {author} {\bibfnamefont {K.}~\bibnamefont
  {Hornberger}},\ }\bibfield  {title} {\enquote {\bibinfo {title} {Entangling
  the free motion of a particle pair: An experimental scenario},}\ }\href@noop
  {} {\bibfield  {journal} {\bibinfo  {journal} {Opt. Spectr.}\ }\textbf
  {\bibinfo {volume} {108}},\ \bibinfo {pages} {188--196} (\bibinfo {year}
  {2010}{\natexlab{a}})}\BibitemShut {NoStop}%
\bibitem [{\citenamefont {Stoof}\ \emph {et~al.}(1988)\citenamefont {Stoof},
  \citenamefont {Koelman},\ and\ \citenamefont {Verhaar}}]{Stoof1988spin}%
  \BibitemOpen
  \bibfield  {author} {\bibinfo {author} {\bibfnamefont {H.~T.~C.}\
  \bibnamefont {Stoof}}, \bibinfo {author} {\bibfnamefont {J.~M. V.~A.}\
  \bibnamefont {Koelman}}, \ and\ \bibinfo {author} {\bibfnamefont {B.~J.}\
  \bibnamefont {Verhaar}},\ }\bibfield  {title} {\enquote {\bibinfo {title}
  {Spin-exchange and dipole relaxation rates in atomic hydrogen: Rigorous and
  simplified calculations},}\ }\href@noop {} {\bibfield  {journal} {\bibinfo
  {journal} {Phys. Rev. B}\ }\textbf {\bibinfo {volume} {38}},\ \bibinfo
  {pages} {4688} (\bibinfo {year} {1988})}\BibitemShut {NoStop}%
\bibitem [{\citenamefont {K\"ohler}\ \emph {et~al.}(2006)\citenamefont
  {K\"ohler}, \citenamefont {G\'oral},\ and\ \citenamefont
  {Julienne}}]{Koehler2006production}%
  \BibitemOpen
  \bibfield  {author} {\bibinfo {author} {\bibfnamefont {T.}~\bibnamefont
  {K\"ohler}}, \bibinfo {author} {\bibfnamefont {K.}~\bibnamefont {G\'oral}}, \
  and\ \bibinfo {author} {\bibfnamefont {P.~S.}\ \bibnamefont {Julienne}},\
  }\bibfield  {title} {\enquote {\bibinfo {title} {Production of cold molecules
  via magnetically tunable {F}eshbach resonances},}\ }\href {\doibase
  10.1103/RevModPhys.78.1311} {\bibfield  {journal} {\bibinfo  {journal} {Rev.
  Mod. Phys.}\ }\textbf {\bibinfo {volume} {78}},\ \bibinfo {pages}
  {1311--1361} (\bibinfo {year} {2006})}\BibitemShut {NoStop}%
\bibitem [{\citenamefont {Gneiting}\ and\ \citenamefont
  {Hornberger}(2010{\natexlab{b}})}]{Gneiting2010molecular}%
  \BibitemOpen
  \bibfield  {author} {\bibinfo {author} {\bibfnamefont {C.}~\bibnamefont
  {Gneiting}}\ and\ \bibinfo {author} {\bibfnamefont {K.}~\bibnamefont
  {Hornberger}},\ }\bibfield  {title} {\enquote {\bibinfo {title} {Molecular
  {F}eshbach dissociation as a source for motionally entangled atoms},}\ }\href
  {\doibase 10.1103/PhysRevA.81.013423} {\bibfield  {journal} {\bibinfo
  {journal} {Phys. Rev. A}\ }\textbf {\bibinfo {volume} {81}},\ \bibinfo
  {pages} {013423} (\bibinfo {year} {2010}{\natexlab{b}})}\BibitemShut
  {NoStop}%
\bibitem [{\citenamefont {Nakajima}(1958)}]{Nakajima1959quantum}%
  \BibitemOpen
  \bibfield  {author} {\bibinfo {author} {\bibfnamefont {S.}~\bibnamefont
  {Nakajima}},\ }\bibfield  {title} {\enquote {\bibinfo {title} {On quantum
  theory of transport phenomena: {S}teady diffusion},}\ }\href@noop {}
  {\bibfield  {journal} {\bibinfo  {journal} {Progr. Theor. Phys.}\ }\textbf
  {\bibinfo {volume} {20}},\ \bibinfo {pages} {948} (\bibinfo {year}
  {1958})}\BibitemShut {NoStop}%
\bibitem [{\citenamefont {Zwanzig}(1960)}]{Zwanzig1960ensemble}%
  \BibitemOpen
  \bibfield  {author} {\bibinfo {author} {\bibfnamefont {R.}~\bibnamefont
  {Zwanzig}},\ }\bibfield  {title} {\enquote {\bibinfo {title} {Ensemble method
  in the theory of irreversibility},}\ }\href@noop {} {\bibfield  {journal}
  {\bibinfo  {journal} {J. Chem. Phys.}\ }\textbf {\bibinfo {volume} {33}},\
  \bibinfo {pages} {1338--1341} (\bibinfo {year} {1960})}\BibitemShut {NoStop}%
\bibitem [{\citenamefont {Breuer}\ and\ \citenamefont
  {Petruccione}(2002)}]{Breuer2002theory}%
  \BibitemOpen
  \bibfield  {author} {\bibinfo {author} {\bibfnamefont {H.-P.}\ \bibnamefont
  {Breuer}}\ and\ \bibinfo {author} {\bibfnamefont {F.}~\bibnamefont
  {Petruccione}},\ }\href@noop {} {\emph {\bibinfo {title} {The Theory of Open
  Quantum Systems}}}\ (\bibinfo  {publisher} {Oxford Univ. Press},\ \bibinfo
  {year} {2002})\BibitemShut {NoStop}%
\bibitem [{\citenamefont {Billy}\ \emph {et~al.}(2008)\citenamefont {Billy},
  \citenamefont {Josse}, \citenamefont {Zuo}, \citenamefont {Bernard},
  \citenamefont {Hambrecht}, \citenamefont {Lugan}, \citenamefont
  {Cl{\'e}ment}, \citenamefont {Sanchez-Palencia}, \citenamefont {Bouyer},\
  and\ \citenamefont {Aspect}}]{Billy2008direct}%
  \BibitemOpen
  \bibfield  {author} {\bibinfo {author} {\bibfnamefont {J.}~\bibnamefont
  {Billy}}, \bibinfo {author} {\bibfnamefont {V.}~\bibnamefont {Josse}},
  \bibinfo {author} {\bibfnamefont {Z.}~\bibnamefont {Zuo}}, \bibinfo {author}
  {\bibfnamefont {A.}~\bibnamefont {Bernard}}, \bibinfo {author} {\bibfnamefont
  {B.}~\bibnamefont {Hambrecht}}, \bibinfo {author} {\bibfnamefont
  {P.}~\bibnamefont {Lugan}}, \bibinfo {author} {\bibfnamefont
  {D.}~\bibnamefont {Cl{\'e}ment}}, \bibinfo {author} {\bibfnamefont
  {L.}~\bibnamefont {Sanchez-Palencia}}, \bibinfo {author} {\bibfnamefont
  {P.}~\bibnamefont {Bouyer}}, \ and\ \bibinfo {author} {\bibfnamefont
  {A.}~\bibnamefont {Aspect}},\ }\bibfield  {title} {\enquote {\bibinfo {title}
  {Direct observation of {A}nderson localization of matter waves in a
  controlled disorder},}\ }\href@noop {} {\bibfield  {journal} {\bibinfo
  {journal} {Nature}\ }\textbf {\bibinfo {volume} {453}},\ \bibinfo {pages}
  {891--894} (\bibinfo {year} {2008})}\BibitemShut {NoStop}%
\bibitem [{\citenamefont {M\"uller}\ \emph {et~al.}(2015)\citenamefont
  {M\"uller}, \citenamefont {Richard}, \citenamefont {Volchkov}, \citenamefont
  {Denechaud}, \citenamefont {Bouyer}, \citenamefont {Aspect},\ and\
  \citenamefont {Josse}}]{Mueller2015suppression}%
  \BibitemOpen
  \bibfield  {author} {\bibinfo {author} {\bibfnamefont {K.}~\bibnamefont
  {M\"uller}}, \bibinfo {author} {\bibfnamefont {J.}~\bibnamefont {Richard}},
  \bibinfo {author} {\bibfnamefont {V.~V.}\ \bibnamefont {Volchkov}}, \bibinfo
  {author} {\bibfnamefont {V.}~\bibnamefont {Denechaud}}, \bibinfo {author}
  {\bibfnamefont {P.}~\bibnamefont {Bouyer}}, \bibinfo {author} {\bibfnamefont
  {A.}~\bibnamefont {Aspect}}, \ and\ \bibinfo {author} {\bibfnamefont
  {V.}~\bibnamefont {Josse}},\ }\bibfield  {title} {\enquote {\bibinfo {title}
  {Suppression and revival of weak localization through control of
  time-reversal symmetry},}\ }\href {\doibase 10.1103/PhysRevLett.114.205301}
  {\bibfield  {journal} {\bibinfo  {journal} {Phys. Rev. Lett.}\ }\textbf
  {\bibinfo {volume} {114}},\ \bibinfo {pages} {205301} (\bibinfo {year}
  {2015})}\BibitemShut {NoStop}%
\bibitem [{\citenamefont {Kuhl}\ \emph {et~al.}(2000)\citenamefont {Kuhl},
  \citenamefont {Izrailev}, \citenamefont {Krokhin},\ and\ \citenamefont
  {St{\"o}ckmann}}]{Kuhl2000experimental}%
  \BibitemOpen
  \bibfield  {author} {\bibinfo {author} {\bibfnamefont {U.}~\bibnamefont
  {Kuhl}}, \bibinfo {author} {\bibfnamefont {F.~M.}\ \bibnamefont {Izrailev}},
  \bibinfo {author} {\bibfnamefont {A.~A.}\ \bibnamefont {Krokhin}}, \ and\
  \bibinfo {author} {\bibfnamefont {H.-J.}\ \bibnamefont {St{\"o}ckmann}},\
  }\bibfield  {title} {\enquote {\bibinfo {title} {Experimental observation of
  the mobility edge in a waveguide with correlated disorder},}\ }\href@noop {}
  {\bibfield  {journal} {\bibinfo  {journal} {Appl. Phys. Lett.}\ }\textbf
  {\bibinfo {volume} {77}},\ \bibinfo {pages} {633--635} (\bibinfo {year}
  {2000})}\BibitemShut {NoStop}%
\bibitem [{\citenamefont {Fern\'andez-Mar\'{\i}n}\ \emph
  {et~al.}(2014)\citenamefont {Fern\'andez-Mar\'{\i}n}, \citenamefont
  {M\'endez-Berm\'udez}, \citenamefont {Carbonell}, \citenamefont {Cervera},
  \citenamefont {S\'anchez-Dehesa},\ and\ \citenamefont
  {Gopar}}]{Fernandez2014beyond}%
  \BibitemOpen
  \bibfield  {author} {\bibinfo {author} {\bibfnamefont {A.~A.}\ \bibnamefont
  {Fern\'andez-Mar\'{\i}n}}, \bibinfo {author} {\bibfnamefont {J.~A.}\
  \bibnamefont {M\'endez-Berm\'udez}}, \bibinfo {author} {\bibfnamefont
  {J.}~\bibnamefont {Carbonell}}, \bibinfo {author} {\bibfnamefont
  {F.}~\bibnamefont {Cervera}}, \bibinfo {author} {\bibfnamefont
  {J.}~\bibnamefont {S\'anchez-Dehesa}}, \ and\ \bibinfo {author}
  {\bibfnamefont {V.~A.}\ \bibnamefont {Gopar}},\ }\bibfield  {title} {\enquote
  {\bibinfo {title} {Beyond {A}nderson localization in 1{D}: Anomalous
  localization of microwaves in random waveguides},}\ }\href {\doibase
  10.1103/PhysRevLett.113.233901} {\bibfield  {journal} {\bibinfo  {journal}
  {Phys. Rev. Lett.}\ }\textbf {\bibinfo {volume} {113}},\ \bibinfo {pages}
  {233901} (\bibinfo {year} {2014})}\BibitemShut {NoStop}%
\bibitem [{\citenamefont {Schwartz}\ \emph {et~al.}(2007)\citenamefont
  {Schwartz}, \citenamefont {Bartal}, \citenamefont {Fishman},\ and\
  \citenamefont {Segev}}]{Schwartz2007transport}%
  \BibitemOpen
  \bibfield  {author} {\bibinfo {author} {\bibfnamefont {T.}~\bibnamefont
  {Schwartz}}, \bibinfo {author} {\bibfnamefont {G.}~\bibnamefont {Bartal}},
  \bibinfo {author} {\bibfnamefont {S.}~\bibnamefont {Fishman}}, \ and\
  \bibinfo {author} {\bibfnamefont {M.}~\bibnamefont {Segev}},\ }\bibfield
  {title} {\enquote {\bibinfo {title} {Transport and {A}nderson localization in
  disordered two-dimensional photonic lattices},}\ }\href@noop {} {\bibfield
  {journal} {\bibinfo  {journal} {Nature}\ }\textbf {\bibinfo {volume} {446}},\
  \bibinfo {pages} {52--55} (\bibinfo {year} {2007})}\BibitemShut {NoStop}%
\bibitem [{\citenamefont {Segev}\ \emph {et~al.}(2013)\citenamefont {Segev},
  \citenamefont {Silberberg},\ and\ \citenamefont
  {Christodoulides}}]{Segev2013anderson}%
  \BibitemOpen
  \bibfield  {author} {\bibinfo {author} {\bibfnamefont {M.}~\bibnamefont
  {Segev}}, \bibinfo {author} {\bibfnamefont {Y.}~\bibnamefont {Silberberg}}, \
  and\ \bibinfo {author} {\bibfnamefont {D.~N.}\ \bibnamefont
  {Christodoulides}},\ }\bibfield  {title} {\enquote {\bibinfo {title}
  {Anderson localization of light},}\ }\href@noop {} {\bibfield  {journal}
  {\bibinfo  {journal} {Nature Phot.}\ }\textbf {\bibinfo {volume} {7}},\
  \bibinfo {pages} {197--204} (\bibinfo {year} {2013})}\BibitemShut {NoStop}%
\end{thebibliography}%

\end{document}